# Superconductivity above 500 K in conductors made by bringing n-alkane into contact with graphite

Yasushi Kawashima

Department of Precision Engineering, School of Engineering, Tokai University, Hiratsuka, Kanagawa 259-1292, Japan.

E-mail: kawasima@keyaki.cc.u-tokai.ac.jp

**In 1986, a cuprate superconductor (Ba-La-Cu-O system) having a critical temperature which goes over the BCS limit (~30 K) was discovered and then a cuprate superconductor (Y-Ba-Cu-O system) with a critical temperature higher than 77 K was discovered. Furthermore, a Hg-based cuprate with a critical temperature of 133 K was found. The 133 K is still the highest critical temperature of conventional superconductors under atmospheric pressure.**

**We have shown that materials obtained by bringing n-alkanes into contact with graphite are capable of conducting electricity with almost no energy loss at room temperature. We here report that the sudden jump in resistance showing a phase transition is observed in the materials during heating by two-probe resistance measurement. The measured critical temperatures of the materials consisting of pitch-based graphite fibers and n-alkanes having 7-16 carbon atoms range from 363.08 to 504.24 K and the transition widths range between 0.15 and 3.01 K. We also demonstrate that superconductors with critical temperatures beyond 504 K are obtained by alkanes with 16 or more carbon atoms.**

# 1. Introduction

In 1911 Heike Kamerlingh Onnes discovered that mercury becomes suddenly zero resistance at temperature close to 4K [1]. This is the discovery of the first superconductor. Gold, silver and copper, which are the best conductors at room temperature, do not become superconductive at all. As temperature is lowered, their resistivity decrease gradually, but they never show a sudden drop. Thereafter, superconductivity was observed for pure metals such as Pb, Nb, and V and alloys such as $Nb_3Al$, $Nb_3Sn$, and $V_3Ga$ [2]. The discovery of superconductors with higher critical temperature became the targets of many researchers. It was mysterious for a long time why superconductivity occurred, but the BCS theory which explains superconductivity was established by J. Bardeen, L. Cooper and R. Schrieffer in 1957 [3]. A superconductor with the critical temperature beyond ~30 K was predicted not to exist according to this BCS theory. However, metallic superconductor $MgB_2$ with a critical temperature of 39 K was discovered in 2001 [4]. This temperature is the highest critical temperature of the metallic superconductor up to now. A cuprate superconductor (La-Ba-Cu-O ceramics) with the critical temperature of 35 K which goes over the BCS limit (~30 K) has been discovered by Bednorz and Müller in 1986 [5], and subsequently, in 1987, the 93 K cuprate superconductor (Y-Ba-Cu-O system) having a critical temperature higher than the boiling point of liquid nitrogen (77 K) was found by M. K. Wu and P. W. Chu [6]. A Hg-based cuprate whose critical temperature is 133 K has been found in 1993 [7], but a cuprate superconductor with a critical temperature higher than that has not been found until now. Although iron-based superconductors were discovered in 2006 [8], the highest critical temperature of iron-based superconductors remains at 56 K up to this time [9]. In 2015, it has been reported that hydrogen sulfide exhibits superconductivity at 203 K under a high pressure of 150 Gpa [10]. This critical temperature is the highest critical temperature currently obtained in conventional superconductors. However, the high pressures exceeding 100 GPa cannot be generated without using a diamond anvil high pressure cell of which the sample chamber is extremely small. On the other hand, Kopelevich et al. reported ferromagnetic and superconducting-like magnetization hysteresis loops in some HOPG samples below and above room temperature suggesting the local superconductivity in graphite in 2000 [11, 12]. Furthermore, in 2012, it was reported that a room temperature superconductor can be made by mixing the water into the graphite particles [13]. However, the authors did not confirm that macroscopic superconducting current passes through this material. Moreover, its critical temperature has not been measured. Recently, Precker1 and Esquinazi et al. identified a transition to zero-resistance state at ~350 K in highly ordered natural graphite sample by resistance measurement, whereas the transition width was ~40 K [14].

We found a possible room-temperature superconductor material obtained by bringing alkanes into contact with the graphite materials [15-18]. We showed that ring current in a ring-shaped container into which n-octane-soaked thin graphite flakes were compressed did not decay for 50 days at room temperature, suggesting that the material is capable of conducting electricity without energy loss at room temperature [15-18]. But the critical temperatures of these materials have not been measured. Therefore, in this study, we attempt to measure the critical temperature of the room temperature superconductor obtained by bringing alkane into contact with graphite material. However, since the above-mentioned material to be measured is an inhomogeneous material, the four-probe method cannot be applied to its resistance measurement. The reason is that there is a possibility that the measurement current path does not necessarily pass through a voltage-measurement terminal in inhomogeneous materials and therefore even if the potential difference between the two voltage-measurement terminals becomes zero, it does not necessarily mean that the resistance becomes zero. Thus, the resistance measurement using four-probe method for inhomogeneous materials causes misunderstanding [19]. The transition from the normal to the superconducting state or from the superconducting to the normal conducting state is accompanied by abrupt change in resistance. Although the result obtained by the resistance measurement using the two-probe method includes contact resistance, sudden jump in resistance at critical temperature can be discriminated by the two-probe method. It has been confirmed that the mixture obtained by bringing the alkane into contact with the graphite material has almost zero resistance at room temperature [15]. If the mixture obtained by bringing alkane into contact with graphite material is gradually heated from room temperature, when it reaches the critical temperature, the resistance of the mixture will jump suddenly. In this study, the critical temperatures of mixtures consisting of graphite materials and alkanes are measured by the two-probe method.

In this research, a pitch-based graphite fiber was used as the graphite material. The sample for critical temperature measurement was prepared by packing the graphite fiber in a polytetrafluoroethylene (PTFE) tube and then injecting alkane into the tube with a syringe. Since the pitch-based graphite fiber is brittle, the fiber is sometimes broken into pieces when packing it in the PTFE tube. Therefore, the resistance of packed pitch-based graphite fibers before injecting alkane into the PTFE tube has a wide range of values. By using the pitch-based graphite fiber, it becomes possible to measure the critical temperature of the sample in which the packed graphite fibers are broken into pieces. In addition, the measurements of the critical temperature are also performed for the samples where alkanes of various carbon numbers are used. On the basis of the experimental results, we will discuss the effect of graphite basal plane surface on superconductivity in the materials.

## 2. Methods

The mixed materials which consist of pitch-based graphite fibers and various alkanes were packed in the PTFE tube shown in Fig. 1 (a). As shown in Fig. 1 (a), the tube has outer diameter of 1.7 mm, an inner diameter of 0.9 mm, a length of 32 mm, and at both ends of the tube, two internal screws of M1 and pitch 0.25 are cut. First, pitch-based graphite fibers (Nippon Graphite Fiber Co., Ltd., XN-100-25Z, no sizing agent, average fiber diameter: 10μm, average fiber length: 25 mm, true density: 2.2 g, resistivity: $1.5 \times 10^{-4}$ Ωcm) were packed in the PTFE tube using metal wire. Secondly, one alkane was chosen from n-heptane, n-octane, n-nonane, n-decane, n-dodecane, n-tridecane and n-hexadecane. Only the alkane was injected into the graphite fibers packed in the PTFE tube by syringe. The experiments were done by using all the alkanes mentioned above. Finally, as shown in Fig. 1 (b), it was capped by screwing two external screws of M1.2, pitch 0.25 mm, a length of 6mm which are made of free-cutting copper (tellurium copper, Cu-0.4 ~ 0.6wt% Te, resistivity: $1.86 \times 10^{-6}$ Ωcm). Weights of graphite fibers packed in the tube range from 0.013 g to 0.018 g. The bulk density of the graphite material packed in the PTFE tube is calculated from the volume of the container ($\phi$ 0.09 × 2 cm) (see Fig. 1(b)), which ranges from 1.02 g / cm$^3$ to 1.41 g / cm$^3$. A copper wire of diameter 0.4 mm is brazed by solder of a high-melting point (the melting point: 586 K) to the heads of the two external screws. In this report, a sample is defined as consisting of the mixture of the graphite fibers and n-alkane packed in the PTFE tube, the two external screws (see Fig. 1 (b)), and the two copper wires brazed to the heads of the screws. Resistance measurement of the sample was performed through the two cooper lead wires. Excluding the case of hexadecane, the resistances of the samples at room temperature ranged from 0.912 Ω to 1.98 Ω, and in the case of hexadecane the resistances of samples were 5.326 Ω, 8.575 Ω and 21.373 Ω. Since the combined resistance of the two copper lead wires and the two screw was 0.0136 Ω, and besides the mixture of graphite material and n-alkane packed in the PTFE tube should have zero resistance at room temperature on the basis of the previous experiments [15], it is considered that the resistance of the sample at room temperature is almost due to contact resistance between the graphite fibers soaked with n-alkane and the two tellurium copper screws.

AC resistance of the sample was measured using LCR meter (E4980A, Keysight Technologies) with Kelvin clip leads (16089B, Keysight Technologies). Measuring frequency was 1 kHz, and a measurement electric current was 10 mA. In order to heat the PTFE tube in which the mixture of the graphite fibers and n-alkane is inserted (see Fig. 1 (b)), the PTFE tube was immersed in n-heptadecane (boiling point: 574 K) which was put in a glass vessel placed on a hot plate (see Fig. 2). As shown in Fig. 2, during heating the vessel by the hot plate, the n-heptadecane was agitated by brass blades installed in the motor to make the temperature of n-heptadecane in the vessel uniform. The tip of K type thermocouple used to measure the temperature of the mixture of the graphite fibers and alkane was brought into contact with the PTFE tube. The K type thermocouple was connected to a digital multimeter (34470A, Keysigh Technologies) through cold junctions (see Fig. 2). The LCR meter and the digital multimeter were linked to a computer by GPIB cables. The alternating-current resistance and thermal data measured by the LCR meter and the digital multimeter were written in the computer memory at intervals of one second.

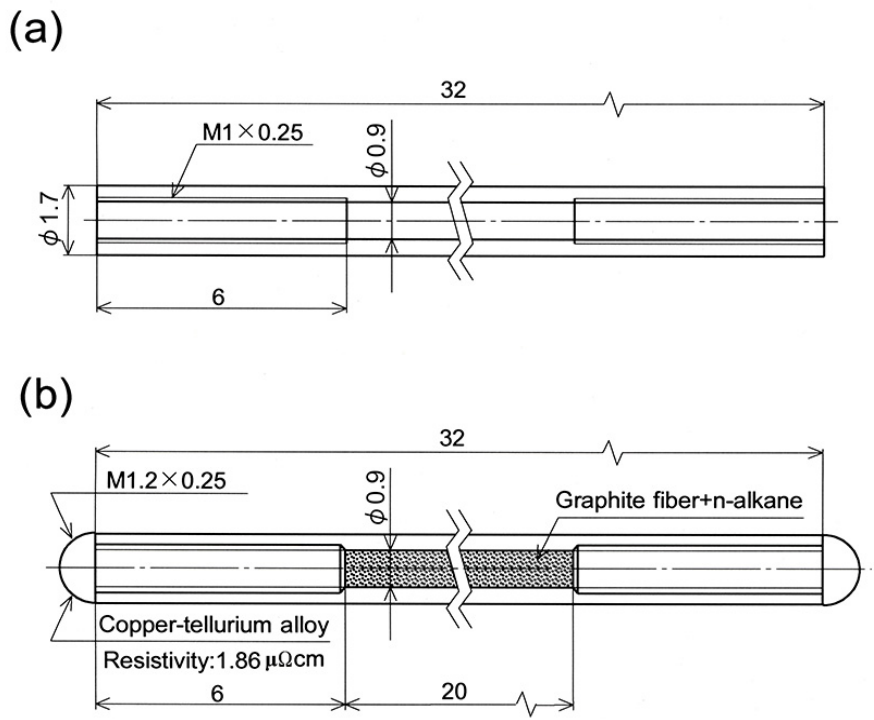

**Figure 1 PTFE tube container for packing the sample and PTFE tube into which graphite fiber and alkane are packed by using two external screws.** (a) Two internal screws of M1 and pitch 0.25 are cut at both ends of the PTFE tube. (b) A mixture of graphite fiber and alkane was crammed into the PTFE tube (a) by using two external screws of M1.2 and pitch 0.25.

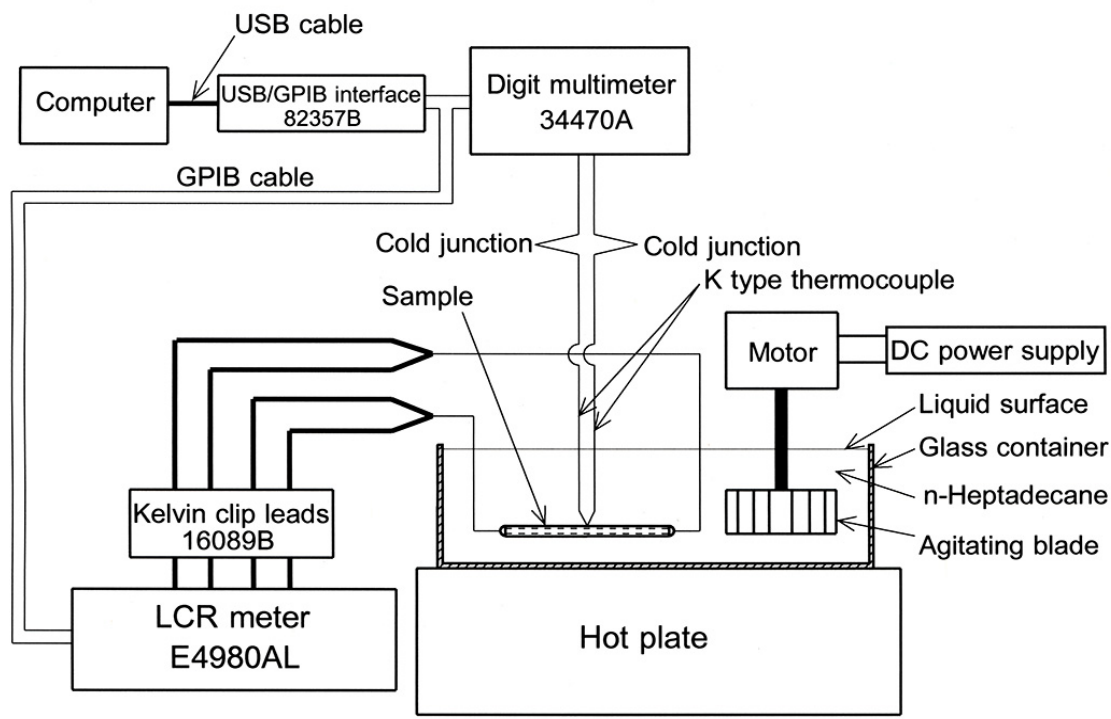

**Figure 2 Schematic of a device for measuring a relationship between the AC resistance and temperature of the graphite fibers and alkane mixture packed in PTFE tube.**

## 3. Results and Discussion

The relationships between the temperature and the alternating-current resistance of the samples, which were obtained by injecting n-heptane, n-octane, n-nonane, n-decane, n-dodecane, n-tridecane and n-hexadecane separately into the graphite fibers packed in the PTFE tube, are shown in Figs. 3-9. All these figures show sudden and complete rise in resistance at temperatures above room temperature, suggesting that the mixtures of graphite fibers and n-alkane in all the samples undergo phase transitions from one state of matter to another. It has been confirmed by the previous experiment [15] that the mixture of graphite fibers and n-alkane should have zero resistance at room temperature. The observed transition shows that the mixtures of graphite fibers and n-alkane were maintained in the superconducting state until the sudden jump in resistance occurs. Therefore, the sudden jump in resistance shows that a phase transition from the superconducting to the normal conducting state occurs in the mixture of graphite fibers and n-alkane.

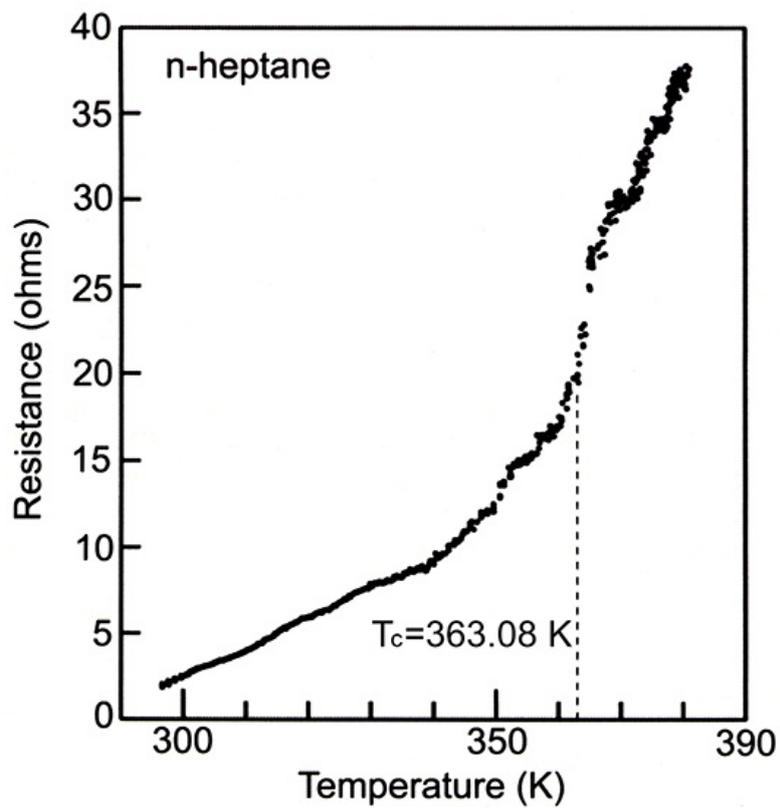

**Figure 3 Resistance versus temperature of a sample consisting of graphite fibers and n-heptane.** Critical temperature: 363.08 K; Amount of change in resistance during the phase transition: 6.501 Ω; the transition width: 1.89 K.

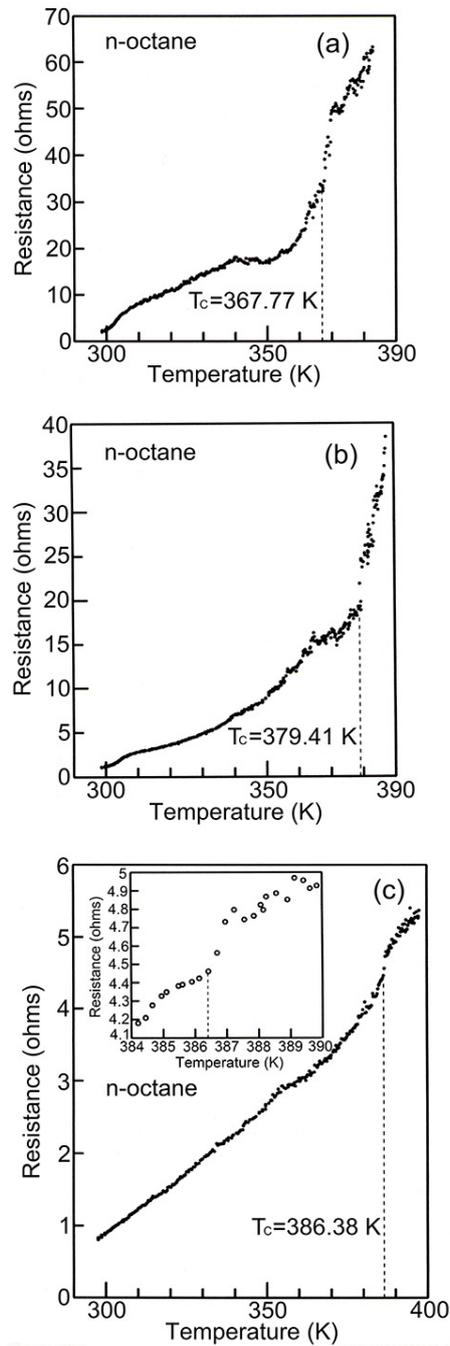

**Figure 4 Resistance versus temperature of a sample consisting of graphite fibers and n-octane.** (a) Critical temperature: 367.77 K; amount of change in resistance during the phase transition: 14.997 Ω; the transition width: 2.39 K. (b) Critical temperature: 379.41 K; amount of change in resistance during the transition: 4.827 Ω; the transition width: 0.15 K. (c) Critical temperature: 386.38 K; amount of change in resistance during the transition: 0.27 Ω; the transition width: 0.54 K. In (c), the inset shows the magnified view of jump in resistance near the critical temperature.

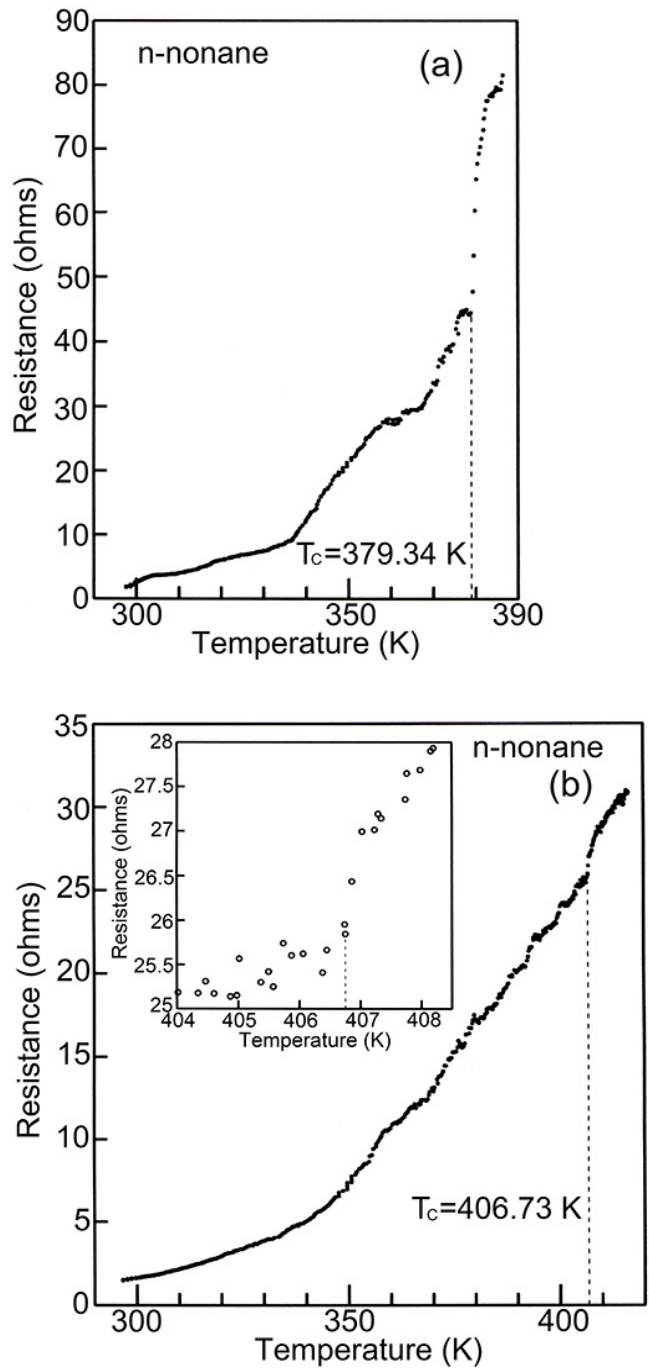

**Figure 5 Resistance versus temperature of a sample consisting of graphite fibers and n-nonane.** (a) Critical temperature: 379.34 K; amount of change in resistance during the phase transition: 23.895 Ω; the transition width: 1.09 K. (b) Critical temperature: 406.73 K; amount of change in resistance during the transition: 1.147 Ω; the transition width: 0.27 K. In (b), the inset shows the magnified view of jump in resistance near the critical temperature.

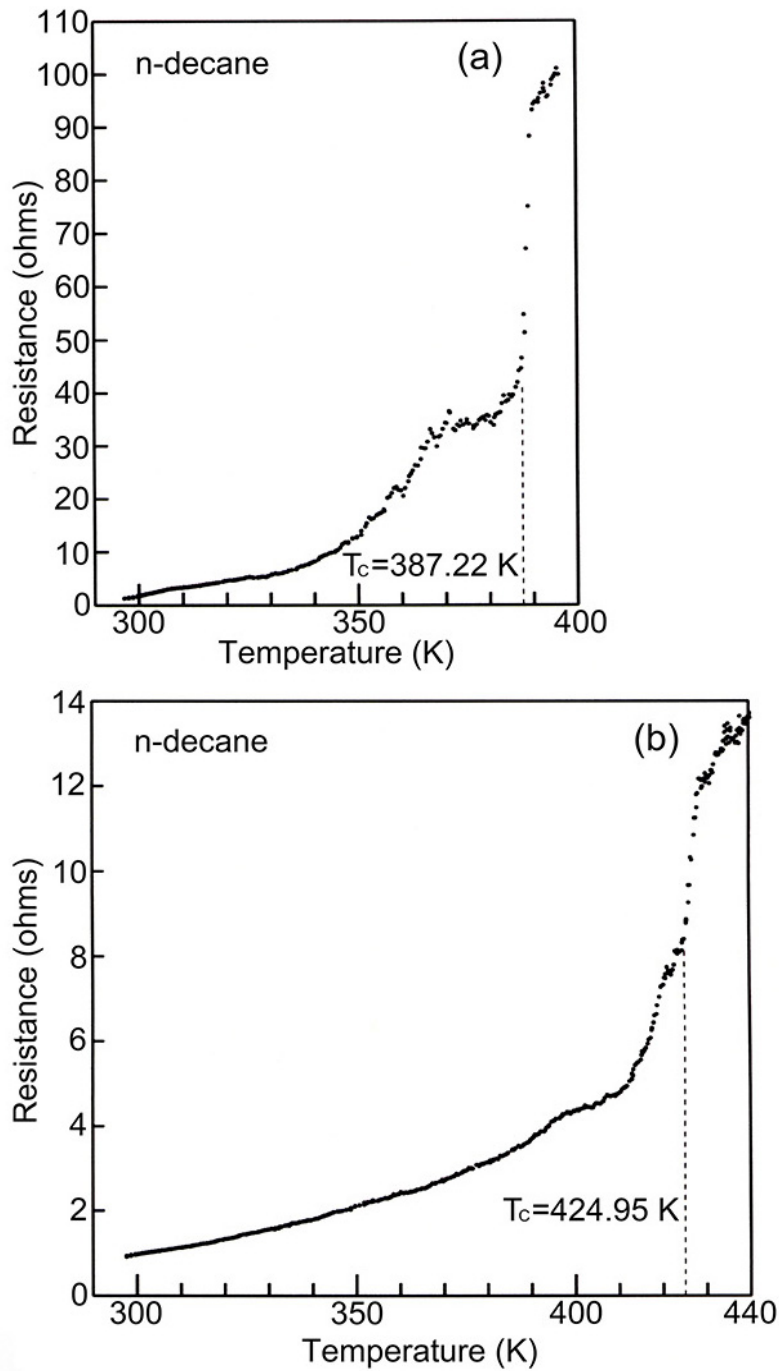

**Figure 6 Resistance versus temperature of a sample consisting of graphite fibers and n-decane.** (a) Critical temperature: 387.22 K; amount of change in resistance during the phase transition: 48.596 Ω; the transition width: 2.83 K. (b) Critical temperature: 424.95 K; amount of change in resistance during the transition: 3.393 Ω; the transition width: 3.01 K.

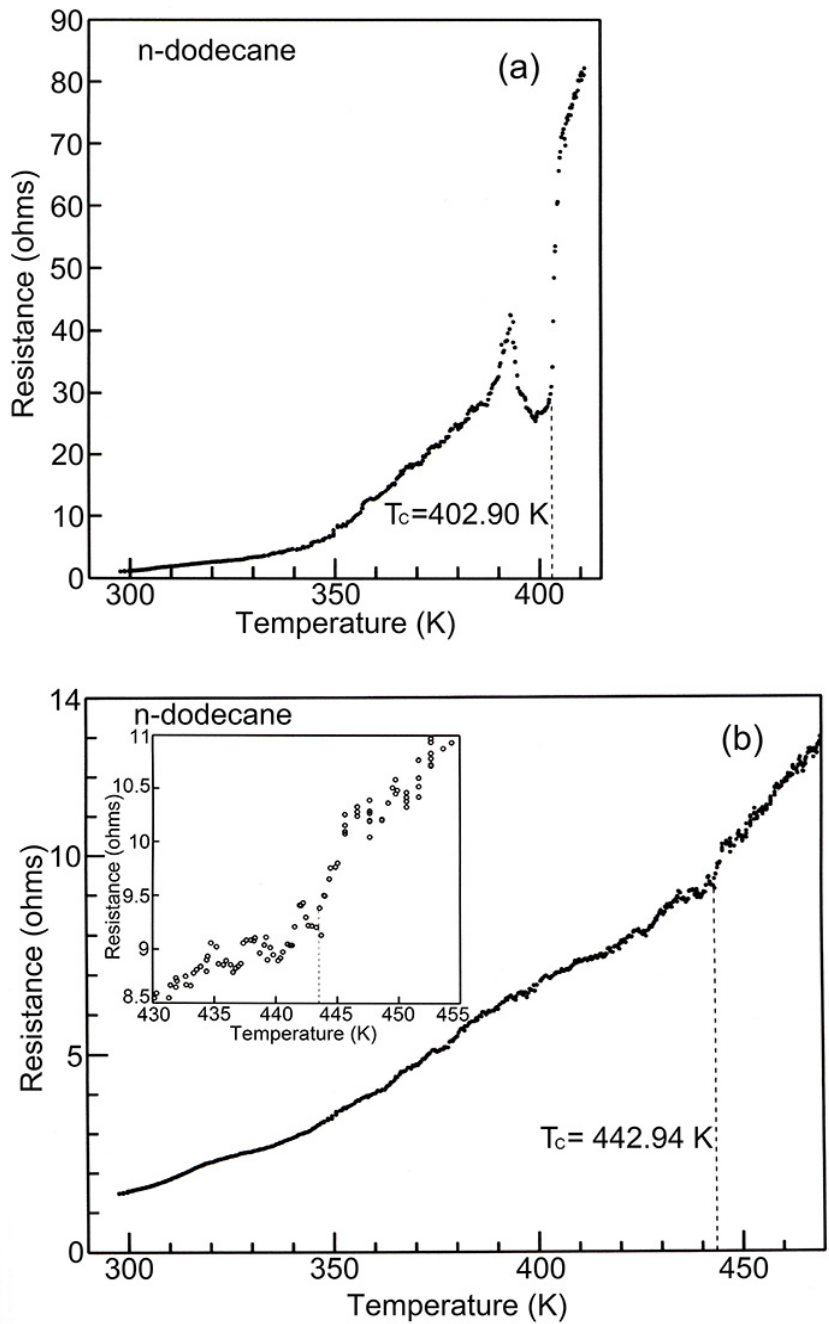

**Figure 7 Resistance versus temperature of a sample consisting of graphite fibers and n-dodecane.** (a) Critical temperature: 402.90 K; amount of change in resistance during the phase transition: 39.984 Ω; the transition width: 2.55 K. (b) Critical temperature: 442.94 K; amount of change in resistance during the transition: 0.856 Ω; the transition width: 2.33 K. In (b), the inset shows the magnified view of jump in resistance near the critical temperature.

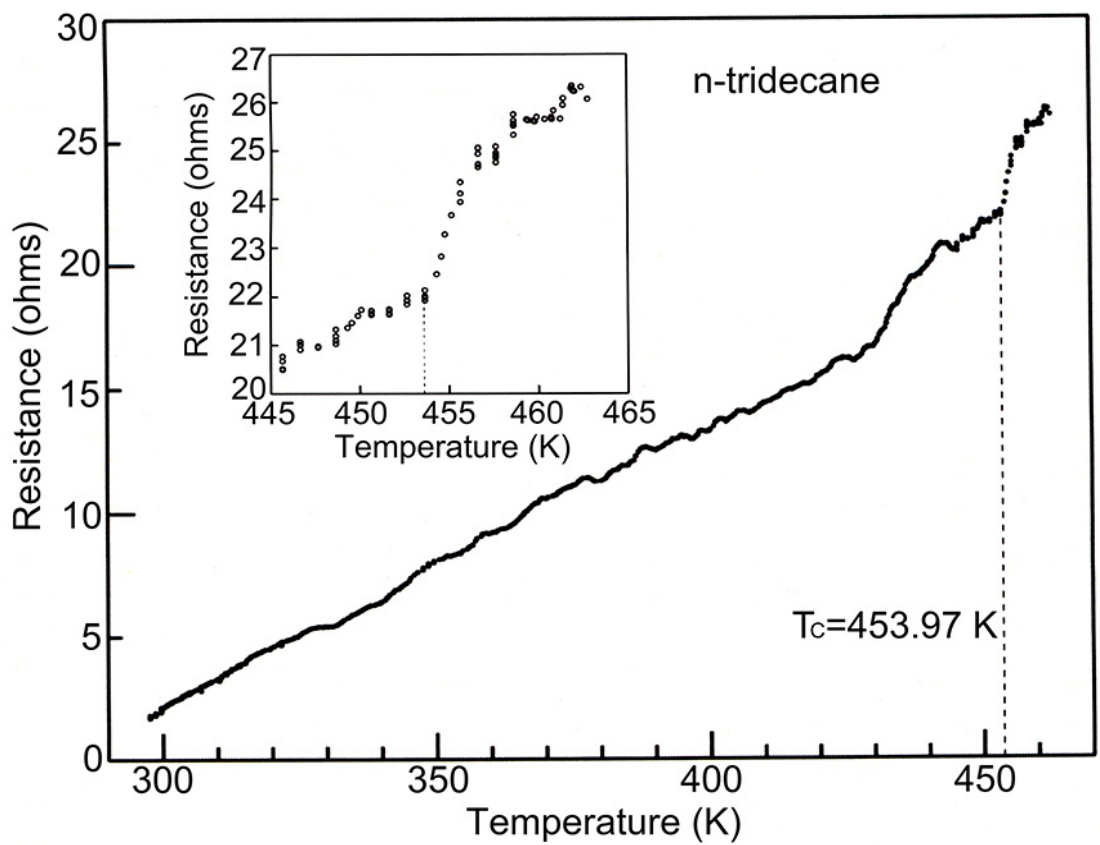

**Figure 8 Resistance versus temperature of a sample consisting of graphite fibers and n-tridecane.** Critical temperature: 453.97 K; amount of change in resistance during the phase transition: 2.553 Ω; the transition width: 1.61 K. The inset shows the magnified view of jump in resistance near the critical temperature.

It can be seen from Figs 3-9 that as the temperature increases, the resistance of the sample, i.e., the contact resistance gradually increases from room temperature. This increase in resistance can be explained as follows. The thermal expansion coefficient of PTFE is of the order of $10^{-4}$ K$^{-1}$ while that of graphite is of the order of $10^{-6}$ K$^{-1}$. Since thermal expansion coefficient of PTFE is approximately two orders of magnitude greater than that of graphite, the PTFE tube expands larger than the graphite materials due to the temperature rise. Therefore, as the temperature increases, the contact pressure decreases and consequently the contact resistance, i.e., the resistance of the sample increases. Thus, the gradual increase in resistance of the samples can be considered to be caused only by the decrease in contact pressure arising from the difference in thermal expansion coefficient between PTFE and graphite. The sudden increase in resistance of the sample shows that the mixture of graphite fibers and n-alkane inserted in the PTFE tube was maintained constant at zero until the phase transition occurs. Accordingly, the temperatures at which the resistances of the samples suddenly rise can be considered to be the critical temperatures.

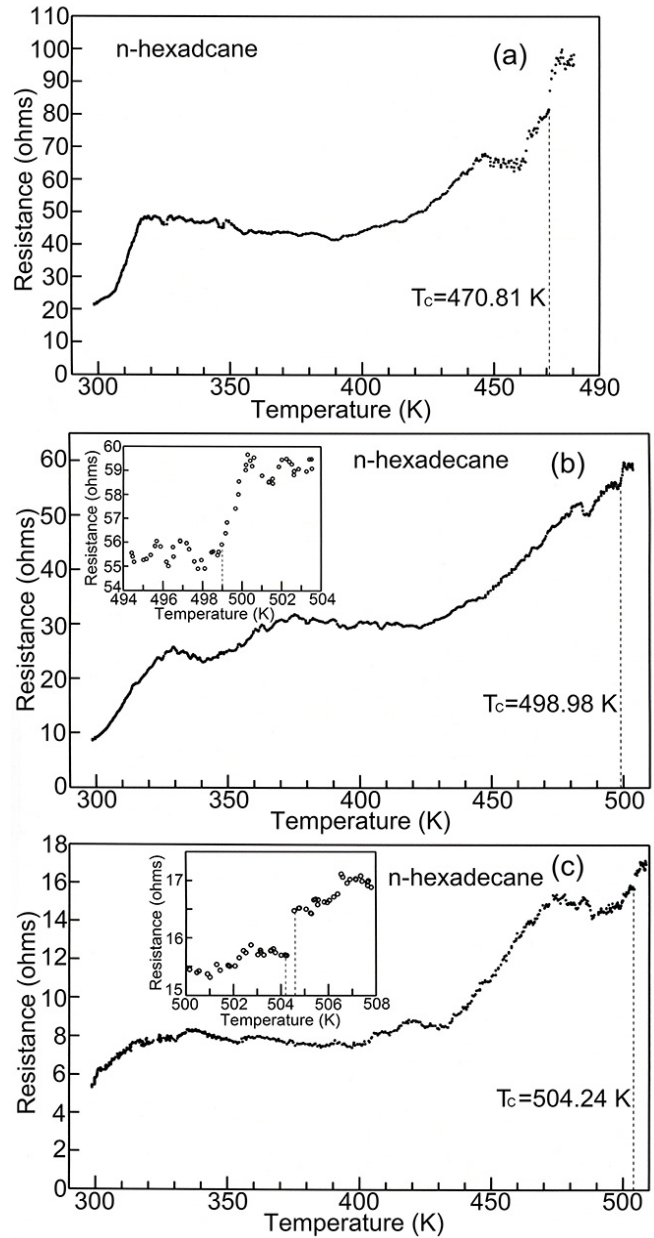

**Figure 9 Resistance versus temperature of a sample consisting of graphite fibers and n-hexadecane.** (a) Critical temperature: 470.81 K; amount of change in resistance during the phase transition: 11.925 Ω; the transition width: 1.19 K. (b) Critical temperature: 498.98 K; amount of change in resistance during the transition: 2.967 Ω; the transition width: 1.2 K. (c) Critical temperature: 504.56 K; amount of change in resistance during the transition: 0.782 Ω; the transition width: 0.32 K. In (b) and (c), insets show the magnified view of jump in resistance near the critical temperatures.

Figure 3 shows the relationship between the temperature and the alternating-current resistance of the sample which were obtained by injecting n-heptane into the graphite fibers packed in the PTFE tube. Figure 3 shows that the resistance suddenly increases from 363.08 K and the sharp rise in resistance is completed at 364.97 K. From this, it can be seen that the resistance of the sample steeply increased by 6.50 Ω with a temperature rise of 1.89 K and the critical temperature is 363.08 K.

Figures 4 (a), (b), and (c) show the relationships between the temperature and the alternating-current resistance of samples which were obtained by injecting n-octane to the graphite fibers packed in the PTFE tube. Figure 4 (a) shows that the resistance suddenly increases by 15.0 Ω while the temperature rises from 367.77 K to 370.16 K, showing that the critical temperature is 367.77 K. Figure 4 (b) shows that the resistance suddenly increases by 4.828 Ω while the temperature rises by 0.15 K from 379.41 K. This suggests that the critical temperature is 379.41 K. Figure 4 (c) shows that while the temperature rises by 0.54 K from 386.38 K, the resistance suddenly increases by 0.27 Ω (see inset in Fig. 4(c)). This suggests that the critical temperature is 386.38 K. The amounts of increase in resistance during the phase transitions obtained from Figs. 4 (a), (b), and (c) are 15.0, 4.828, and 0.27 Ω, respectively.

Figures 5 (a) and (b) show the relationships between the temperature and the alternating-current resistance of samples which were obtained by injecting n-nonane to the graphite fibers packed in the PTFE tube. Figure 5 (a) shows that the resistance makes a sudden jump at 379.34 K and it increases by 23.90 Ω while the temperature rises from 379.34 K to 380.43 K, showing that the critical temperature is 379.34 K. Figure 5 (b) shows that the resistance suddenly increases by 1.15 Ω while the temperature rises by 0.27 K from 406.73 K, showing that the critical temperature is 406.73 K (see inset in Fig. 5(b)). Therefore, it is seen from Figs 5 (a) and (b) that when the critical temperatures are 379.34 K and 406.73 K, the amounts of change in resistance during the phase transition are 23.90 and 1.15 Ω, respectively.

Figures 6 (a) and (b) show the relationships between the temperature and the alternating-current resistance of samples which were obtained by injecting n-decane to the graphite fibers packed in the PTFE tube. Figure 6 (a) shows that the resistance makes a sudden jump at 387.22 K and it increases by 48.60 Ω while the temperature rises from 387.22 K to 390.05 K, showing that the critical temperature is 387.22 K. Figure 6 (b) shows that the resistance steeply increases by 3.39 Ω while the temperature rises by 3.01 K from 424.95 K, showing that the critical temperature is 424.95 K. It is found from Figs. 6 (a) and (b) that when the critical temperatures are 387.22 K and 424.95 K, the amounts of increase in resistance during the transition are 48.60 Ω and 3.39 Ω, respectively.

Figures 7 (a) and (b) show the relationships between the temperature and the alternating-current resistance of the samples which were obtained by injecting n-dodecane into the graphite fibers packed in the PTFE tube. Figure 7 (a) shows that the resistance makes a sudden jump at 402.90 K and it increases by 39.98 Ω while the temperature rises from 402.90 K to 405.445 K, showing that the critical temperature is 402.90 K. Figure 7 (b) shows that while the temperature rises 2.33 K from 442.94 K, the resistance suddenly increases by 0.856 Ω, suggesting that the critical temperature is 442.94 K (see inset in Fig.7(b)). It can be seen from Figs 7 (a) and (b) that when the critical temperatures are 402.90 K and 442.94, the amounts of increase in resistance during the transition are 39.98 Ω and 0.856 Ω, respectively.

Figure 8 shows the relationship between the temperature and the alternating-current resistance of a sample which was obtained by injecting n-tridecane to the graphite fibers crammed in the PTFE tube. Figure 8 shows that while the temperature rises from 453.97 K to 455.58 K, the resistance steeply increases by 2.553 Ω, showing that the critical temperature is 453.97 K and furthermore the amount of increase in resistance during the transition is 2.553 Ω (see inset in Fig. 8).

Figures 9 (a), (b), and (c) show the relationships between the temperature and the alternating-current resistance of samples which were obtained by injecting n-hexadecane into the graphite fibers packed in the PTFE tube. Figure 9 (a) shows that the resistance increases suddenly by 11.92 Ω while the temperature increases from 470.81 K to 472.00 K, showing that the critical temperature is 470.81 K. Figure 9 (b) shows that while the temperature rises by 1.2 K from 498.98 K, the resistance steeply increases 2.97 Ω (see inset in Fig. 9(b)). This shows that the critical temperature is 498.98 K. Fig. 9 (c) shows that the resistance makes a sudden jump at 504.24 K (see inset in Fig. 9 (c)) and it increases by 0.78 Ω due to a temperature rise of 0.32 K from 504.24 K. This suggests that the critical temperature is 504.24 K. Figures 9 (a), (b) and (c) show that when the critical temperatures are 470.81 K, 498.98 K and 504.24 K, the amounts of increase in resistance of the sample during the transition are 11.925, 2.967 and 0.782 Ω, respectively.

It can be seen from Figs. 4-7 and Fig. 9 that as the amount of increase in resistance during the phase transition decreases, the critical temperature becomes higher. The mixture of graphite fibers and n-alkane inserted in the PTFE tube is considered to have the zero-resistance until the resistance suddenly jumps up. Therefore, the amount of change in resistance during the phase transition is considered to be nearly equal to the resistance of the sample where only the graphite fibers are packed before injecting n-alkane into the PTFE tube. Accordingly, it is found that the smaller the resistance of the graphite fibers packed in the PTFE tube before injecting alkane into the tube, the higher the critical temperature.

Figure 10 shows the relationships between the critical temperature and the amount of increase in resistance during the transition in the cases of using n-octane and n-hexadecane. Figure 10 indicates that there exists a nearly linear relationship between the critical temperature and the amount of increase in resistance during the phase transition.

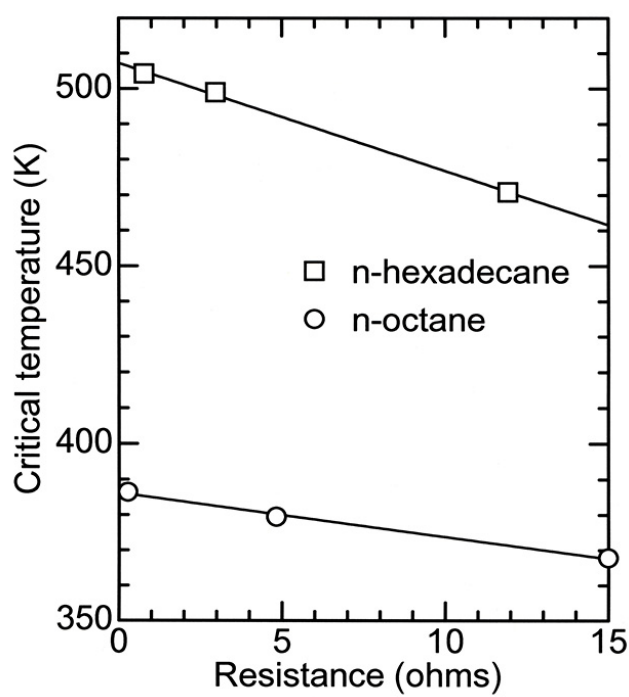

**Figure 10 Relationships between the critical temperature and the amount of change in resistance during the transition in the cases of using n-octane and n-hexadecane.** Straight lines in the figure were obtained by the least squares method.

Figure 11 shows the relationship between the carbon number (n) of alkanes and the critical temperature ($T_c$). In Fig. 11, these critical temperatures are values when the amount of increase in resistance during the phase transition is 2.553 Ω. As can be seen from Fig. 10, since it can be assumed that there exists a near-linear relationship between the critical temperature and the amount of change in resistance during the phase transition, the critical temperatures at 2.553 Ω in the case of alkanes with 8-10, 12, 16 carbon atoms were obtained by linear interpolation. Figure 11 shows that as the carbon number of n-alkane increases, the critical temperature becomes higher.

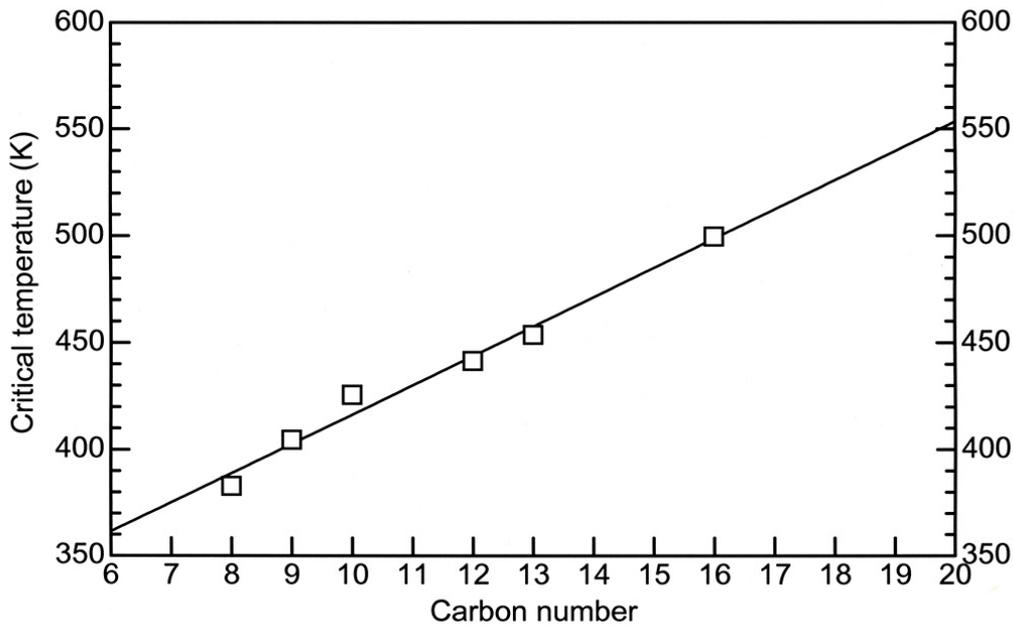

**Figure 11 Relationship between the critical temperature and the carbon number of alkanes.** A line is drawn to make the relationship easier to see.

The resistances of the graphite fibers packed in the PTFE tube before injecting alkanes into the tube, which are obtained from the amount of change in resistance during the phase transition, range from 0.27 to 48.6 Ω. Since the dimensions of the sample container of PTFE tube is the diameter of 0.09 cm, length 2 cm, the range of apparent resistivity of the sample calculated from the above resistances was between $1.54 \times 10^{-1}$ Ωcm and $8.59 \times 10^{-4}$ Ωcm. The resistivity of the graphite fiber used is $1.5 \times 10^{-4}$ Ωcm. Even taking into account the bulk density of the graphite material packed in the PTFE tube (1.02~1.41 g / cm$^3$), the apparent resistivity of $1.54 \times 10^{-1}$ Ωcm is considered to be too high. The pitch-based carbon fiber used is very brittle because of its highly developed graphitizability [20]. Therefore, the high resistance of the samples is attributable to the breakage of the graphite fibers packed in the TPFE tube. If the graphite fibers break down into shorter fibers, the resistance of the sample will become higher accordingly. The surface of graphite fiber almost consists of basal plane surface. However, it can be considered that in the case of high resistance, the graphite fibers crack finely and therefore the ratio of basal surface to edge surface is reduced. Thus, it can be deduced that as the ratio of basal surface to edge surface becomes smaller, the critical temperature becomes lower. This suggests that basal surface plays an important role in the superconductivity in the mixture of graphite and alkane.

We believe that in the superconductors obtained by bringing alkanes into contact with graphite material, superconductivity can be ascribable to protons abstracted by the graphite basal surface from n-alkanes [21]. We appreciate that all protons causing superconductivity should move coherently without activation energy on the basal surface [16-18, 21]. As the ratio of basal plane surface to edge surface becomes smaller, the proton is more easily separated from the basal surface by the influence of heat, so that the superconducting state collapses. Therefore, the smaller the ratio of basal surface to edge surface, the lower the critical temperature.

Figure 11 shows that the greater the number of carbon atoms of alkane, the higher the critical temperature. The boiling points of n-alkanes increase regularly with the increase in the number of carbon atoms. Furthermore, as the temperature approaches the boiling point of n-alkane, the thermal motion of the n-alkane molecules grows more intense. It can be considered that the superconducting state in the mixture of graphite materials and alkane may be broken by the thermal motion of alkane molecules. Figure 11 indicates that when n-alkanes with 16 or more carbon atoms are used, the critical temperature exceed 500 K.

## 4. Conclusions

We have observed a sharp rise in resistance of mixture of graphite fibers and alkanes inserted in the PTFE tube during heating them, using the two-probe-resistance measurements. This observation shows that a phase transition from the superconducting to the normal conducting state occurred in the mixture of graphite fibers and n-alkane. In other words, this indicates that the mixture of graphite fibers and n-alkane packed in the PTFE tube remains superconductive until its resistance rapidly rises, and the temperature at which the resistance suddenly rises is the critical temperature. We have noticed that when the pitch-based graphite fibers are packed in the PTFE tube, they may be broken into pieces depending on how they are packed into the tube. The finer the graphite fibers break, the greater the resistance of the sample before injecting alkane into the PTFE tube packed with the graphite fiber. From the relationship between the amount of change in resistance at the phase transition and the critical temperature, we found that the critical temperature decreases as the fiber is finely divided. That is, the critical temperature decreases, as the ratio of the basal plane surface to the edge plane surface decreases. This fact suggests that the basal plane plays an important role in superconductivity. Furthermore, we have found that the greater the carbon number of alkane, that is, the higher the boiling point of alkane, the higher the critical temperature. We have demonstrated that superconductors having critical temperatures exceeding 500 K can be obtained by using n-alkanes having 16 or more carbon atoms.

## Acknowledgements

The author would like to thank Prof. Theodore H Geballe for his on-target instruction.